\documentclass[aps,prd,twocolumn,amsmath,amssymb,floatfix,nofootinbib,superscriptaddress]{revtex4-1}
\usepackage[utf8]{inputenc}
\usepackage{xcolor}
\usepackage{amsmath}
\usepackage{amssymb}
\usepackage{graphicx}
\usepackage{natbib}
\usepackage{hyperref}
\usepackage{aas_macros}
\usepackage[normalem]{ulem}
\usepackage{siunitx}

\newcommand{\nhat}[1][]{\ensuremath{\hat{\mathbf{n}}}}
\newcommand{\obsmap}[1]{\ensuremath{#1^{\mathrm{obs}}}}
\newcommand{\lenmap}[1]{\ensuremath{#1^{\mathrm{len}}}}
\newcommand{\unlmap}[1]{\ensuremath{#1^{\mathrm{prim}}}}
\newcommand{\rotmap}[1]{\ensuremath{#1^{\mathrm{rot}}}}
\newcommand{\biasmap}[1]{\ensuremath{#1^{\mathrm{biased}}}}
\newcommand{\vect}[1]{\ensuremath{\boldsymbol{#1}}}
\newcommand{\vl}{\vect{\ell}}
\newcommand{\vL}{\vect{L}}

\newcommand{\smu}{Department of Physics,
Southern Methodist University, 3215 Daniel Ave, Dallas, Texas 75275, USA}

\begin{document}

\title{Deep Learning for Primordial $B$-mode Extraction}
\author{Eric~Guzman}
\affiliation{\smu}
\author{Joel~Meyers}
\affiliation{\smu}
\date{\today}

\begin{abstract}
The search for primordial gravitational waves is a central goal of cosmic microwave background (CMB) surveys.
Isolating the characteristic $B$-mode polarization signal sourced by primordial gravitational waves is challenging for several reasons: the amplitude of the signal is inherently small; astrophysical foregrounds produce $B$-mode polarization contaminating the signal; and secondary $B$-mode polarization fluctuations are produced via the conversion of $E$ modes.
Current and future low-noise, multi-frequency observations enable sufficient precision to address the first two of these challenges such that secondary $B$ modes will become the bottleneck for improved constraints on the amplitude of primordial gravitational waves.
The dominant source of secondary $B$-mode polarization is gravitational lensing by large scale structure.
Various strategies have been developed to estimate the lensing deflection and to reverse its effects the CMB, thus reducing confusion from lensing $B$ modes in the search for primordial gravitational waves.
However, a few complications remain.
First, there may be additional sources of secondary $B$-mode polarization, for example from patchy reionization or from cosmic polarization rotation.
Second, the statistics of delensed CMB maps can become complicated and non-Gaussian, especially when advanced lensing reconstruction techniques are applied.
We previously demonstrated how a deep learning network, ResUNet-CMB, can provide nearly optimal simultaneous estimates of multiple sources of secondary $B$-mode polarization.
In this paper, we show how deep learning can be applied to estimate and remove multiple sources of secondary $B$-mode polarization, and we further show how this technique can be used in a likelihood analysis to produce nearly optimal, unbiased estimates of the amplitude of primordial gravitational waves.

\end{abstract}

\maketitle


\section{Introduction}
\label{sec:Intro}

Current and next-generation cosmic microwave background (CMB) surveys will map temperature and polarization anisotropies with unprecedented precision~\cite{SimonsObservatory:2018koc, SimonsObservatory:2025wwn, Abazajian:2016yjj, SPT-3G:2014dbx, Hui:2018cvg, Li:2017drr, Sehgal:2019ewc, NASAPICO:2019thw, LiteBIRD:2022cnt}. The forthcoming data will enable greatly increased precision on the measurement of cosmological parameters with broad implications for cosmology and fundamental physics. Among the key targets for current and upcoming CMB surveys is the search for primordial gravitational waves~\cite{Planck:2006aa, BICEP2:2018kqh, Abazajian:2016yjj, SimonsObservatory:2018koc, Li:2017drr, NASAPICO:2019thw}.

Cosmic inflation generically predicts the production of primordial gravitational waves, with an amplitude directly related to the expansion rate during inflation~\cite{Starobinsky:1979ty, Rubakov:1982df, Fabbri:1983us, Abbott:1984fp}.  Detection of primordial gravitational waves would provide very strong evidence in favor of inflation, and it would also give direct insight into the energy scale at which inflation took place.

Primordial gravitational waves leave a characteristic signature in the CMB in the form of $B$-mode polarization.  Primordial $B$-mode polarization is  generated by primordial tensor fluctuations but not by scalar density fluctuations at first order in perturbation theory~\cite{Kamionkowski:1997na, Seljak:1996gy}. Detection of the primordial $B$-mode polarization thus acts as a 'smoking gun' for primordial gravitational waves, and in turn for an inflationary period in the early universe.

Measurement of the primordial $B$-mode polarization is an especially difficult task for at least three distinct reasons: the expected cosmological signal (such as that predicted from models of inflation) is intrinsically weak; astrophysical foregrounds including the polarized emission from Galactic dust act as a source of confusion for the primordial signal; and non-primordial $B$-mode polarization is present due to various effects that convert $E$-mode polarization into $B$-mode polarization, adding another source of confusion for the signal of interest; see Ref.~\cite{Kamionkowski:2015yta} for a review.  Future observations searching for primordial gravitational waves therefore need to have very low instrumental noise on degree angular scales where the cosmological signal peaks, they require sufficient frequency coverage to disentangle astrophysical foregrounds, and precise observations on small angular scales are required to reconstruct and remove the effects that produce secondary $B$ modes~\cite{Kamionkowski:2015yta, CMB-S4:2020lpa}.

We focus in this paper on secondary CMB anisotropies, which act as a source of confusion in the search for primordial gravitational waves.
Effects that distort the primary anisotropies induce non-stationary statistics of the observed CMB sky. Local changes to CMB statistics allow the fields responsible for distorting the CMB to be reconstructed from the induced correlation between anisotropies on different angular scales, which would otherwise be independent~\cite{Hu:2001kj, Yadav:2009za}. A key component to placing tighter constraints on primordial gravitational waves is our ability to disentangle the $B$-mode polarization resulting from secondary anisotropies and thereby reduce the variance of the observed $B$-mode polarization~\cite{Knox:2002pe, Kesden:2002ku}.

The most prominent source of secondary CMB anisotropies that has been measured arises from gravitational lensing by large scale structure.
Gravitational lensing of the CMB refers to the deflection of CMB photons due to variations in the gravitational potential intervening between the surface of last scattering and our telescopes~\cite{Lewis:2006fu}.  The effects of CMB lensing have been observed at high significance in both temperature and polarization~\cite{Planck:2018lbu, ACT:2023dou, ACT:2023kun, SPT:2023jql, SPT-3G:2024atg}.  Reconstructing and reversing the effects of CMB lensing, a process known as CMB delensing~\cite{Seljak:2003pn, Smith:2010gu,Green:2016cjr, Hotinli:2021umk}, has been shown to reduce the observed $B$-mode power~\cite{SPT:2017ddy} and to improve constraints on the amplitude of primordial gravitational waves~\cite{SPTpol:2020rqg}. 

Gravitational lensing is not the only source of secondary $B$-mode polarization.
The process by which the first stars and galaxies reionized the universe was inhomogeneous.  As a result, the line-of-sight free electron density, and therefore the CMB optical depth to reionization, is expected to vary across the sky.  This variation leads to screening and scattering of CMB photons, which produces $B$-mode polarization, even in the absence of primordial gravitational waves~\cite{Dvorkin:2009ah}.  Much like gravitational lensing, the mode coupling induced by patchy screening can be used to reconstruct fluctuations in the spatially varying optical depth~\cite{Dvorkin:2008tf}, and this effect can then be reversed, a process known as demodulation. In addition to the screening of polarization sourced around recombination, the inhomogeneous density of free electrons around reionization can produce $B$-mode polarization via scattering of the incoming temperature quadrupole~\cite{Dvorkin:2008tf,Dvorkin:2009ah}.   The effects of patchy screening and scattering on polarization have not yet been observed in the CMB~\cite{Namikawa:2017uke, ACT:2024jdf}, but they must be present at some level given what is known about the physics of reionization.

In addition to gravitational lensing and patchy reionization, other effects may produce secondary $B$-mode polarization.  Cosmic polarization rotation, whereby the plane of linear polarization is rotated as photons propagate through the Universe, can convert $E$-mode polarization to $B$-mode polarization.  Cosmic polarization rotation can result from new parity violating physics, like the electromagnetic coupling of an axion-like field that evolves over cosmic time~\cite{Carroll:1989vb, Harari:1992ea, Carroll:1998zi, Lue:1998mq,Marsh:2015xka}.  It can also be induced by Faraday rotation in the presence of primordial or astrophysical magnetic fields~\cite{Kosowsky:1996yc}.  Cosmic polarization rotation can be uniform across the sky, or it can be anisotropic.  Like the effects described above, anisotropic cosmic polarization rotation results in non-stationary CMB statistics, allowing reconstruction of the rotation angle via the induced off-diagonal mode coupling~\cite{Kamionkowski:2008fp,Yadav:2009eb,Gluscevic:2009mm}.  The effects of cosmic polarization rotation can also be reversed through the process of derotation.

Reversing the effects of lensing (delensing), cosmic polarization rotation (derotation), and patchy reionization (demodulation) all reduce the amplitude of secondary $B$-mode polarization.  Each process thereby enables improved sensitivity in the search for primordial gravitational waves (at least to the degree that the associated secondary $B$ modes act as a source of confusion.)

The tensor-to-scalar ratio, $r$, is defined to be the ratio of primordial gravitational wave power to that of density fluctuations. 
The current best upper limit on the tensor-to scalar ratio is $r<0.036$ at $95\%$ confidence which is provided by the Background Imaging of Cosmic Extragalactic Polarization (BICEP) and Keck collaboration~\cite{BICEP:2021xfz}.  The precision of the best current measurements are limited by instrumental noise and astrophysical foregrounds. Upcoming CMB surveys, like those expected from Simons Observatory~\cite{Ade:2018sbj}, 
LiteBIRD~\cite{LiteBIRD:2022cnt}, and PICO~\cite{Hanany:2019lle}, will either detect primordial gravitational waves or improve upper limits on $r$ by more than an order of magnitude.  Reaching this precision will require that secondary anisotropies are reconstructed and removed, since the power of the $B$ modes from gravitational lensing is known to exceed the desired level of precision. It may also become important to remove the effects of patchy reionization and cosmic polarization rotation, though neither of these effects has yet been observed in the CMB.

Several approaches are being pursued to delens the CMB, including internal delensing with maximum likelihood methods~\cite{Hirata:2003ka}, maximum a posterioiri techniques~\cite{Carron:2017mqf, Carron:2017vfg, Carron:2018lcr, Legrand:2021qdu, Legrand:2023jne}, and Bayesian methods~\cite{Millea:2017fyd, Millea:2020cpw}.  There are also techniques based on estimates of the lensing field from external tracers~\cite{Smith:2010gu, Sherwin:2015baa, Larsen:2016wpa, BaleatoLizancos:2021owo}.  Recently machine learning techniques have been developed to estimate CMB lensing and to perform delensing~\cite{Caldeira:2018ojb, Heinrich:2022hsf, Yan:2023oan, Yi:2024vkd}.

In this paper, we present an analysis pipeline designed to simultaneously reconstruct multiple effects that produce secondary $B$-mode polarization using a trained ResUNet-CMB network, reverse those effects to reduce the secondary $B$-mode polarization amplitude, and subsequently estimate the tensor-to-scalar ratio $r$.  
We will demonstrate that this pipeline provides a robust procedure for using the ResUNet-CMB estimates to reverse the effects that lead to secondary anisotropies from both lensing and cosmic polarization rotation. Additionally, we will use Bayesian inference to measure the tensor-to-scalar ratio after delensing and derotation of the $Q$ and $U$ polarization maps.

This paper is organized as follows.  
In Section~\ref{sec:SecondaryAnisotropies} we describe how secondary CMB anisotropies are produced from gravitational lensing and cosmic polarization rotation, and we describe how these effects are estimated and reversed within the context of quadratic estimators.
We describe our deep learning network ResUNet-CMB, the simulated data pipeline, and network training  in Section~\ref{sec:DeepLearning}.
In Section~\ref{sec:Results}, we show how ResUNet-CMB can be employed to mitigate secondary CMB anisotropies and provided improved estimates of the primordial gravitational wave amplitude.
We conclude in Section~\ref{sec:Conclusion}.

\section{Secondary Anisotropies}
\label{sec:SecondaryAnisotropies}

Weak gravitational lensing of the CMB occurs when the photons traveling towards our telescopes are gravitationally deflected by cosmological structure along their path. The photons we receive appear to originate from a position that differs by a small angle from their true origin. In real space, this deflection alters the $T$, $Q$, and $U$ maps of the CMB as
\begin{align}
    \label{eqn:LensedFields}
    \lenmap{T}(\nhat) &= \unlmap{T}(\nhat + \nabla \phi (\nhat)) \, , \nonumber\\
    (\lenmap{Q} \pm i\lenmap{U})(\nhat) &= (\unlmap{Q} \pm i\unlmap{U})(\nhat + \nabla \phi (\nhat)) \, ,
\end{align}
where $\nhat$ is the line-of-sight direction, $\phi$ is the CMB lensing potential, the `prim' superscript identifies the primordial CMB maps, and the `len' superscript denotes the lensed maps. Importantly, gravitational lensing deflection converts $E$-mode polarization to $B$-mode polarization, leading to a contribution to the observed $B$-mode power at leading order in the lensing power of the form
\begin{align}
    C_{\ell}^{BB,\mathrm{len}} =& \int{\frac{d^2 \vect{\ell}_{1}}{(2 \pi)^2 }} \left[ (\vect{\ell}_1 \cdot \vect{\ell}_2) \sin{2\varphi_{\vect{\ell}_1,\vect{\ell}}} \right]^2   C_{\ell_2}^{\phi\phi} C_{\ell_1}^{EE}  \, ,
    \label{eq:LensedBB}
\end{align}
where $\vl_2 = \vl-\vl_1$.

The deflection of fluctuations leads to non-stationary statistics of observed maps giving off-diagonal mode coupling in harmonic space.
Using the flat-sky approximation, we can express the mode coupling induced by gravitational lensing as~\cite{Hu:2001kj}
\begin{equation}
    \langle x(\vl_1) x'(\vl_2) \rangle_\mathrm{CMB} = f_\beta^{\phi}(\vl_1, \vl_2)\phi(\vL) \, ,
    \label{eq:lensing_mode_coupling}
\end{equation}
where $x,x'\in \{T, E, B\}$, $\vL = \vl_1 + \vl_2 \neq 0$, and $\beta$ denotes a particular pairing of $T$, $E$, and $B$, $\beta\ \in \{TT, TE, TB, EE, EB, BB\}$.
Explicit expressions for $f_\beta^\phi(\vl_1, \vl_2)$ are given in Table~\ref{tab:mode_couplings}.

Cosmic polarization rotation leaves the temperature unchanged but mixes $Q$ and $U$ polarization
\begin{align}
     \label{eqn:AlphaRotation}
     \rotmap{T} &= \unlmap{T} \nonumber\\
     \rotmap{Q}(\nhat) &= \unlmap{Q}(\nhat) \cos{2 \alpha(\nhat)} - \unlmap{U}(\nhat) \sin{2 \alpha(\nhat)} \nonumber\\
     \rotmap{U}(\nhat) &= \unlmap{Q}(\nhat) \sin{2 \alpha(\nhat)} + \unlmap{U}(\nhat) \cos{2 \alpha(\nhat)} \, .
\end{align}
Just as in the case of lensing, cosmic polarization rotation converts $E$-modes to $B$-modes, adding a contribution to the $B$-mode power
\begin{align}
        C_{\ell}^{BB,\mathrm{rot}} =& \int{\frac{d^2 \vect{\ell}_{1}}{(2 \pi)^2 }} \left[ 2 \cos{2\varphi_{\vect{\ell}_1,\vect{\ell}}} \right]^2   C_{\ell_2}^{\alpha\alpha} C_{\ell_1}^{EE}  \, .
    \label{eq:RotationBB}
\end{align}
Also like lensing, polarization rotation leads to mode coupling~\cite{Kamionkowski:2008fp, Namikawa:2016fcz}, which we describe with a function $f_\beta^\alpha(\vl_1,\vl_2)$, shown for each polarization combination in Table~\ref{tab:mode_couplings}.

\begin{table*}[tb!]
\begin{tabular}{l|c|c}
$\beta$ & $f_\beta^{\phi}(\vl_1, \vl_2)$  & $f_\beta^{\alpha}(\vl_1, \vl_2)$ \\ \hline
$TT$    &  $C_{\ell_1}^{TT}(\vL \cdot \vl_1) + C_{\ell_2}^{TT}(\vL \cdot \vl_2)$                   &      $0$                              \\
$TE$    & $C_{\ell_1}^{TE} \cos \varphi_{\vl_1 \vl_2} (\vL \cdot \vl_1) + C_{\ell_2}^{TE}(\vL \cdot \vl_2)$            &                  $-2 C_{\ell_1}^{TE} \sin 2 \varphi_{\vl_1, \vl_2}$                              \\
$TB$    &  $C_{\ell_1}^{TE} \sin 2\varphi_{\vl_1 \vl_2} (\vL \cdot \vl_1)     $                    &    $2 C_{\ell_1}^{TE} \cos 2 \varphi_{\vl_1, \vl_2}$                              \\
$EE$    & $ \left[C_{\ell_1}^{EE}(\vL \cdot \vl_1) + C_{\ell_2}^{EE}(\vL \cdot \vl_2)\right] \cos 2\varphi_{\vl_1 \vl_2}$    &  $-2\left[ C_{\ell_1}^{EE} + C_{\ell_2}^{EE} \right] \sin 2 \varphi_{\vl_1, \vl_2}$                                \\
$EB$    &   $\left[C_{\ell_1}^{EE}(\vL \cdot \vl_1) - C_{\ell_2}^{BB}(\vL \cdot \vl_2)\right] \sin 2\varphi_{\vl_1 \vl_2} $           &  $2 \left[C_{\ell_1}^{EE} - C_{\ell_2}^{BB} \right] \cos 2 \varphi_{\vl_1, \vl_2}$    \\
$BB$    &   $\left[C_{\ell_1}^{BB}(\vL \cdot \vl_1) + C_{\ell_2}^{BB}(\vL \cdot \vl_2)\right] \cos 2\varphi_{\vl_1 \vl_2}$         &  $-2\left[ C_{\ell_1}^{BB} + C_{\ell_2}^{BB} \right] \sin 2 \varphi_{\vl_1, \vl_2}$                     
\end{tabular}
\label{tab:mode_couplings}
\caption{Mode coupling functions for each polarization combination induced by gravitational lensing ($f_\beta^{\phi}(\vl_1, \vl_2)$) and by cosmic polarization rotation ($f_\beta^{\alpha}(\vl_1, \vl_2)$).}
\end{table*}

The off-diagonal mode coupling induced by gravitational lensing and cosmic polarization rotation (as well as other effects) allows them to be reconstructed with a quadratic estimator~\cite{Hu:2001kj}.  We can construct an estimator for the field $\Gamma \in \{\phi, \alpha\}$ as a weighted product of two CMB maps
\begin{equation}
    \hat{\Gamma}(\vL) = A^\Gamma_\beta(\vL) \int \frac{d^2 \ell_1}{(2\pi)^2} x(\vl_1) x'(\vl_2) F_\beta^\Gamma(\vl_1,\vl_2) \, ,
    \label{eq:QE_generic}
\end{equation}
where $\vl_2 = \vL-\vl_1$.  The normalization $A_\beta^\Gamma(\vL)$ is chosen to ensure that $\hat{\Gamma}$ provides an unbiased estimate of $\Gamma$ and is given by
\begin{equation}
    A_\beta^\Gamma(\vL) = \left[\int \frac{d^2 \ell_1}{(2\pi)^2} f_\beta^\Gamma(\vl_1,\vl_2) F_\beta^\Gamma(\vl_1,\vl_2)\right]^{-1}
    \label{eq:QE_norm} \, .
\end{equation}
The filter $F_\beta^\Gamma(\vl_1,\vl_2)$ is chosen to minimize the variance of the reconstructed field and is given by
\begin{equation}
    F_\beta^{\Gamma}(\vl_1, \vl_2) = \frac{ C_{\ell_1}^{x'x'}C_{\ell_2}^{xx} f_\beta^\Gamma(\vl_1,\vl_2) - C_{\ell_1}^{xx'}C_{\ell_2}^{xx'} f_\beta^\Gamma(\vl_2,\vl_1) }{C_{\ell_1}^{xx}C_{\ell_2}^{x'x'}C_{\ell_1}^{x'x'}C_{\ell_2}^{xx} - (C_{\ell_1}^{xx'}C_{\ell_2}^{xx'})^2} \, ,
    \label{eq:optimal_filter}
\end{equation}
where power spectra in this expression refer to the observed spectra, including instrumental noise power.  With this choice of filter, the variance of the reconstructed map is
\begin{align}
    \label{eqn:GeneralReconNoise}
    \left\langle \hat{\Gamma}_\beta(\vL) \hat{\Gamma}_{\beta'}(\vL') \right\rangle = (2\pi)^2 \delta(\vL + \vL') \left(C_\ell^{\Gamma\Gamma} + N_{\beta\beta'}^{\Gamma\Gamma}(L) \right) \, ,
\end{align}
with $N_{\beta\beta}^{\Gamma\Gamma}(L) = A_\beta^\Gamma(\vL)$.

For the remainder of the paper,  observed CMB maps (\obsmap{Q}, \obsmap{U}, \obsmap{T}) refer to those which are first rotated according to Eq.~\eqref{eqn:AlphaRotation}, lensed with Eq.~\eqref{eqn:LensedFields}, then have a noise map added with a power spectrum according to $N_\ell = \Delta^2 \exp\left(\ell(\ell+1)\frac{\theta_\mathrm{FWHM}^2}{8\log2}\right)$, where we take $\Delta_P = \sqrt{2}\Delta_T$. The final observed maps in real space are defined as
\begin{align}
    \label{eqn:LensedRotatedFields}
    \obsmap{T}(\nhat) &= \lenmap{(\rotmap{( \unlmap{T}(\nhat) )})} + T^N(\nhat) \nonumber\\
    (\obsmap{Q} \pm i\obsmap{U})(\nhat)  &= \lenmap{(\rotmap{( (\unlmap{Q} \pm i\unlmap{U})(\nhat) )})} \nonumber \\
    &\qquad+Q^N(\nhat) \pm iU^N(\nhat)\, .
\end{align}

The additional $B$-mode power present due to $E$-to-$B$ conversion from gravitational lensing and polarization rotation poses a challenge for searches for primordial gravitational waves.  First, the observed $B$-mode power is larger in the presence of these effects than one would expect for any value of the tensor-to-scalar ratio, which would lead to a biased inference of $r$ if not accounted for.  However, this is not the main problem with the secondary $B$ modes, since it is straightforward to subtract off the expected contribution to the observed $B$-mode power if we know the spectra involved.  These contributions are given for lensing and rotation in Eqs.~\eqref{eq:LensedBB} and \eqref{eq:RotationBB}.  The larger challenge comes from the fact that simply subtracting the power does not remove the additional variance contributed by the secondary anisotropies in the observed $B$-mode maps.  In order to mitigate this extra variance, we need to remove the secondary $B$-modes at map level.  It is for this reason that the map-level reconstruction of the fields that produce the secondaries is essential in searches for primordial gravitational waves.

We would therefore like to achieve a low-variance estimate of any fields that convert $E$ modes to $B$ modes, such that we can remove the secondary $B$ modes at map level. In general, the variance of a map reconstructed with the quadratic estimator depends on the total observed CMB power.  The total observed power includes contributions from the variance of the primary CMB, the instrumental noise power (and astrophysical foregrounds), and the secondary anisotropies from all sources of statistical anisotropy including lensing and polarization rotation.  At low noise levels, where the secondary $B$ modes make the dominant contribution to the observed $B$-mode power, this motivates going beyond the quadratic estimator for reconstruction.

As discussed in Section~\ref{sec:Intro}, several methods for lensing reconstruction and delensing beyond the quadratic estimator are being pursued.  Here our goal is to show that the simultaneous reconstruction of lensing and cosmic polarization rotation provided by the ResUNet-CMB network, along with a delensing and derotation pipeline based on those estimates, is capable of producing nearly optimal estimates of the tensor-to-scalar ratio.  In order to establish a basis of comparison, we need to specify what is meant by optimal performance.  The procedure we describe here is similar to the one we utilized in Ref.~\cite{Guzman:2021ygf} to analyze the performance of the ResUNet-CMB polarization rotation reconstruction.

It has been shown in Ref.~\cite{Smith:2010gu} that an iterated estimate of the lensing reconstruction noise and the residual $B$-mode power after delensing produces results that match closely with what is expected from maximum likelihood reconstruction~\cite{Hirata:2003ka}. Here we follow a similar procedure, applied to both lensing and polarization rotation.  Specifically, starting with an observed $B$-mode power that includes the contributions from lensing, Eq.~\eqref{eq:LensedBB}, and polarization rotation, Eq.~\eqref{eq:RotationBB}, we estimate the lensing reconstruction noise via Eq.~\eqref{eq:QE_norm}.  We then estimate the contribution to the $B$-mode power that would remain in the map after delensing using the reconstructed lensing map $\hat{\phi}$.  Specifically, one estimates the amount of $B$-mode power that can be removed by calculating the lensed $B$-mode power that would arise from the Wiener-filtered lensing map  applied to the Wiener-filtered $E$-mode polarization map, which is then subtracted from the lensed $B$-mode power. This results in a residual lensing $B$-mode power after delensing of
\begin{align}
    C_{\ell}^{BB,\mathrm{len},\mathrm{res}} =& \int{\frac{d^2 \vect{\ell}_{1}}{(2 \pi)^2 }} \left[ (\vect{\ell}_1 \cdot \vect{\ell}_2) \sin{2\varphi_{\vect{\ell}_1,\vect{\ell}}} \right]^2 \nonumber \\
    & \times C_{\ell_2}^{\phi\phi} C_{\ell_1}^{EE} \left[ 1 - \frac{C_{\ell_2}^{\phi\phi}}{C_{\ell_2}^{\phi\phi,\mathrm{obs}}} \frac{C_{\ell_1}^{EE}}{C_{\ell_1}^{EE,\mathrm{obs}}} \right] \, .
    \label{eq:LensingResidualBB}
\end{align}
Next, we estimate the polarization rotation reconstruction noise using Eq.~\eqref{eq:QE_norm}, where now the observed $B$-mode power has been reduced due to the replacement of $C_\ell^{BB,\mathrm{len}}$ with $C_{\ell}^{BB,\mathrm{len},\mathrm{res}}$.  Then, we estimate how much $B$-mode power remains after derotation by the reconstructed rotation map $\hat{\alpha}$,
\begin{align}
        C_{\ell}^{BB,\mathrm{rot},\mathrm{res}} =& \int{\frac{d^2 \vect{\ell}_{1}}{(2 \pi)^2 }} \left[2  \cos{2\varphi_{\vect{\ell}_1,\vect{\ell}}} \right]^2 \nonumber \\
    & \times  C_{\ell_2}^{\alpha\alpha} C_{\ell_1}^{EE} \left[ 1 - \frac{C_{\ell_2}^{\alpha\alpha}}{C_{\ell_2}^{\alpha\alpha,\mathrm{obs}}} \frac{C_{\ell_1}^{EE}}{C_{\ell_1}^{EE,\mathrm{obs}}} \right] \, .
    \label{eqRotationResidualBB}
\end{align}
We can then obtain an improved estimate of the lensing field due to the reduction of $C_\ell^{BB,\mathrm{rot}}$ to $C_{\ell}^{BB,\mathrm{rot},\mathrm{res}}$.  This whole procedure is then iterated to convergence.  The resulting reconstruction noise and residual $B$-mode power serve as our estimate of what could be obtained in an optimal delensing and derotation analysis.

This iterative procedure treated at the level of the power spectra approximates the effects of lensing and rotation as independent, though these effects are not independent in real data nor in our simulations.  Nonetheless, it provides a useful baseline against which we can compare the outputs of the ResUNet-CMB predictions without the complications of attempting a true map-level iterative reconstruction scheme.

Even though we do not  implement an iterative map-level procedure for delensing and derotation here, we will make a few comments that will be relevant for the map-level implementation of our procedure based on ResUNet-CMB reconstruction discussed below.  
As mentioned above, when noise is present in the polarization maps, we need to filter the maps before carrying out the delensing and derotation.  Define a Wiener filter for each field as
\begin{equation}
    W_\ell = \frac{C_\ell}{C_\ell^{\mathrm{obs}}} \, ,
    \label{eq:Wiener_Filter}
\end{equation}
then the process of delensing and derotation can be achieved as
\begin{align}
    &(Q\pm iU)^\mathrm{delen,derot}(\nhat) = \left[\bar{W}\star (Q\pm iU)^\mathrm{obs}\right](\nhat) \nonumber \\
    & \quad + \left(\left(\left[W\star (Q\pm iU)^\mathrm{obs}\right](\nhat)\right)^{\mathrm{delen}}_{W\star\hat{\phi}}\right)^{\mathrm{derot}}_{W\star\hat{\alpha}}
    \label{eq:Delensed_Derotated_Map}
\end{align}
where $\bar{W}_\ell = 1-W_\ell$.  The second term of this expression is constructed by starting with the observed polarization map, applying a polarization Wiener filter, applying inverse lensing with the filtered estimated lensing field $W\star\hat{\phi}$, then applying inverse rotation with the filtered estimated polarization rotation field $W\star\hat{\alpha}$.  The first term comes from the observed polarization map filtered by the complement of the Wiener filter $\bar{W}_\ell$.  This first term is required to conserve the total power of the delensed and derotated map while not applying the delensing and derotation operations to the noisy modes~\cite{Green:2016cjr}.  For the simulated maps we consider here, the $E$-mode polarization is signal-dominated across all relevant scales and the Wiener filtering of the polarization has little effect, though in general, the Wiener filtering is important to avoid biasing the residual $B$-mode power, especially in cases with strongly scale-dependent polarization noise or foreground power.  Note that this procedure describes an implementation of all-orders delensing and derotation, in contrast to template delensing and derotation.  The latter refers to estimating the induced $B$-mode map and simply subtracting it from the observed map, rather than transforming the observed fields.  For a comparison of all-orders and template delensing, see Ref.~\cite{BaleatoLizancos:2020jby}.


\begin{figure*}[t!]
    \centering
    \includegraphics[width=0.95\textwidth]{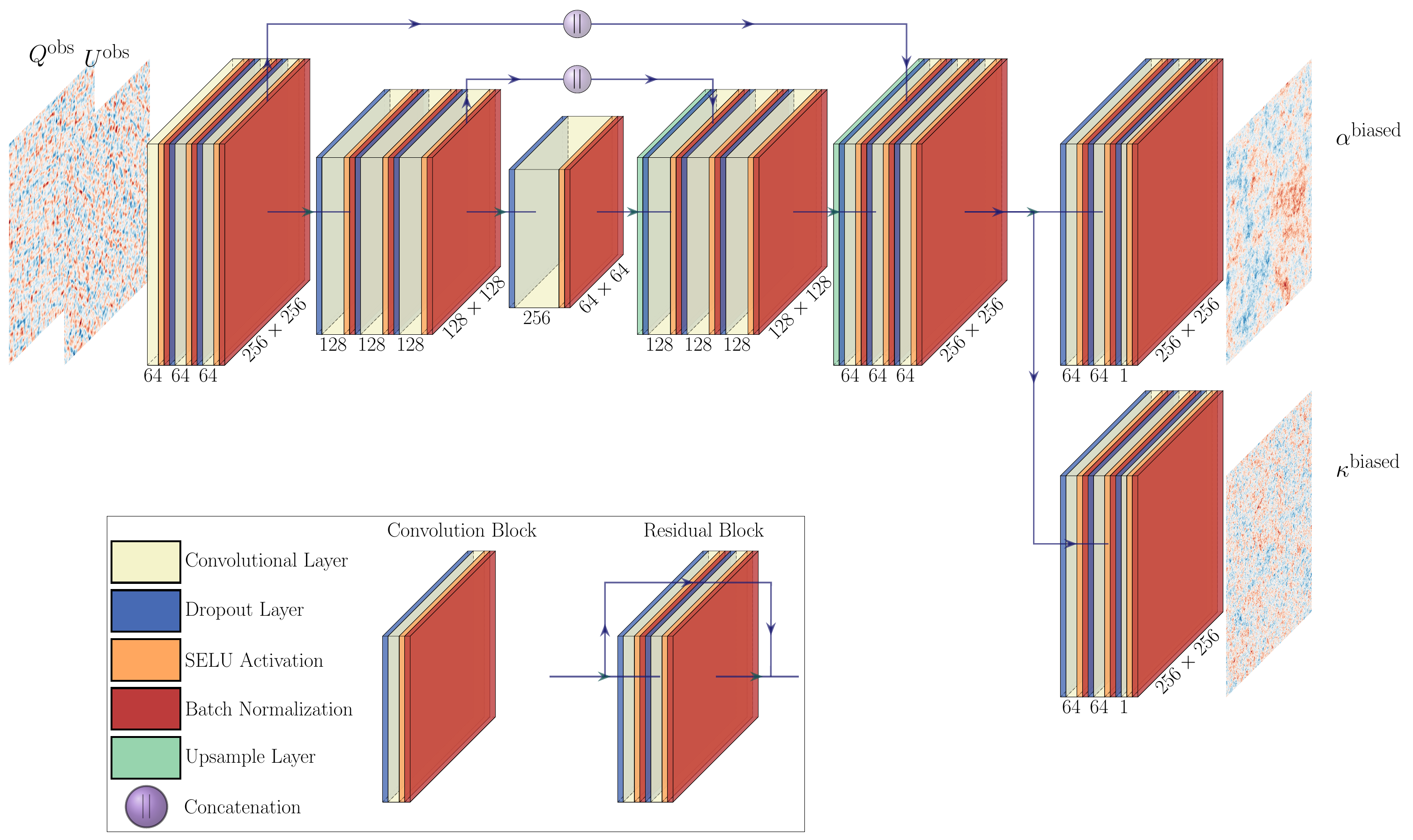}
    \caption{
    Modified ResUNet-CMB architecture with residual connections excluded for clarity. The network input tensor has a shape of $(256, 256, 2)$ which consists of the stacked ($\obsmap{Q}$, $\obsmap{U}$) CMB polarization maps. There are only two outputs in this architecture, ($\biasmap{\alpha}$, $\biasmap{\kappa}$) which come from the final batch normalization layers in their respective branches. The numbers on the side of each convolutional layer are the number of filters used for that layer. The numbers at the bottom of the front of the output batch normalization layers are the image size. Down-sampling by using a stride of \num{2} is done in the encoder phase instead of using pooling layers.
    (The graphic was made with publicly available code from \url{https://github.com/HarisIqbal88/PlotNeuralNet}.)}
    \label{fig:FP_ResUNet-Structure}
\end{figure*}

\section{ResUNet-CMB Architecture, Data Pipeline, and Training}\label{sec:DeepLearning}

Over the past several years, computer vision techniques have found widespread application to problems in cosmology; for recent reviews of machine learning applications in this field, see Refs.~\cite{Dvorkin:2022pwo,Rojas:2025wdm}. One of the prominent tools for computer vision tasks is the convolutional neural network (CNN). CNNs are a type of artificial neural network designed to extract information and features from images. Cosmological observations, which can be represented as two- or three-dimensional images, are prime targets for CNN applications. 

In this paper, we focus on the application of a specific type of CNN, the Residual U-Net (ResUNet). In particular, we choose to use the ResUNet-CMB network architecture originally defined in Ref.~\cite{Guzman:2021nfk} and extended in Ref.~\cite{Guzman:2021ygf}; ResUNet-CMB is similar in structure to the network described in Ref.~\cite{Caldeira:2018ojb}. In our previous works, we showed that the ResUNet architecture was well suited for the simultaneous reconstruction of fields that distort our observations of the CMB. The modular nature of the network made extension and modification for cosmological problems straightforward. In this paper, we will use the ResUNet-CMB network and modify it in order to detect primordial gravitational waves by reducing the secondary anisotropies by derotating and delensing the modified observed CMB maps.

In this section, we will focus on highlighting the changes made to the ResUNet-CMB model from Refs.~\cite{Guzman:2021nfk, Guzman:2021ygf} in order to use the network in an analysis pipeline that determines the tensor-to-scalar ratio, $r$.

\subsection{ResUNet-CMB}
The modified ResUNet-CMB network\footnote{The code for the ResUNet-CMB architecture and the updated data pipeline is located at \url{https://github.com/EEmGuzman/resunet-cmb}.} is shown in Fig.~\ref{fig:FP_ResUNet-Structure}. Similar to what was done in Refs.~\cite{Guzman:2021nfk, Guzman:2021ygf}, implementation of the network was done with the Keras~\cite{chollet2015keras} package of TensorFlow~\cite{tensorflow2015-whitepaper}. While we continue to use $Q$ and $U$ polarization maps as input to the network, the images have been simulated to be $256 \times 256$ pixels instead of the $128\times128$ pixels used in previous work~\cite{Guzman:2021nfk,Guzman:2021ygf,Caldeira:2018ojb}.

More information regarding the specifics of the maps will be described below in Sec.~\ref{sec:DataPipeNetTraining}. The $Q$ and $U$ maps are concatenated along their color channels forming a tensor of shape $(256, 256, 2)$ and are fed into the network. In this modified ResUNet-CMB network, only two output branches are present with one for $\alpha$ and the other for $\kappa$. Both $\alpha$ and $\kappa$ also now have a dimension size of $256 \times 256$ pixels. Despite the increase in number of pixels per image, all intermediate convolution blocks and residual blocks have the same core structure as was utilized in Refs.~\cite{Guzman:2021nfk,Guzman:2021ygf}. There has also been no change to the number of filters, strides, or any other layer parameter set in the original network  design. Additionally, since all map sizes were increased, the minimum representation layer -- where the network transitions from encoder to decoder -- now has an image size of $64\times64$ pixels instead of $32\times32$ pixels. Despite the increase in image size, the number of trainable network parameters remains around \num{5000000} since the layer parameters were not modified. All residual and skip connection patterns also remain the same as previous iterations of the ResUNet-CMB network.

No dropout layer is used before the first convolutional layer of the network in order to avoid the irreversible loss of information. All residual connections in the network follow the same structure as in Ref.~\cite{Guzman:2021nfk}. The input layer is added element wise to the output of the second convolution block, the output of the second convolutional block is added to that of the fourth, and so on. When residual connections connect convolution blocks with different numbers of filters, a convolutional layer with a linear activation is added into the residual connection that modifies the image to the size of the target layer. We also include a batch normalization layer after each residual connection convolutional layer. The residual connections help with the backward propagation of the gradients during training which allows the network to easily learn the identity function. Images of the same size in the encoder are concatenated along the color channel with the matching image sizes in the decoder. These concatenations give rise to the U-Net shape of the architecture~\cite{ronneberger2015unet}. These encoder-decoder skip connections mirror those seen in Ref.~\cite{Guzman:2021nfk}.

\subsection{Data pipeline}
\label{sec:DataPipeNetTraining}

To simulate the CMB maps we first generate theory spectra with \texttt{CAMB}\footnote{\url{https://camb.info}}~\cite{Lewis:1999bs,Howlett:2012mh} from a fiducial cosmology defined by the parameters: $H_{0} = 67.4$~km~s$^{-1}$~Mpc$^{-1}$, $\Omega_{b}h^{2} = 0.0224$, $\Omega_{c}h^{2} = 0.120$, $n_{s} = 0.966$, $\bar{\tau} = 0.0561$, and $A_{s} = 2.10\times 10^{-9}$, consistent with Planck 2018 cosmology~\cite{Planck:2018vyg}. Anisotropic polarization rotation is included in our simulations, so we also use the same scale-invariant power spectrum included in Ref.~\cite{Guzman:2021ygf} which is of the form $\ell^2 C_\ell^{\alpha\alpha}/(2\pi) = 0.014$~deg$^2$.

From the theory spectra, we simulate $256 \times 256$ two-dimensional maps each covering a $10^{\circ} \times 10^{\circ}$ flat patch of the sky. We first generate a set of CMB maps $(\unlmap{T}, \unlmap{E}, \unlmap{B})$, convert them to $(\unlmap{Q}, \unlmap{U})$, and then simulate distortion fields $(\alpha, \kappa)$ using modified code from CMB-S4\footnote{\url{https://github.com/jeffmcm1977/CMBAnalysis_SummerSchool}} and \texttt{Orphics}\footnote{\url{https://github.com/msyriac/orphics}}. To get the observed polarization maps, $(\unlmap{Q}, \unlmap{U})$ are first lensed then rotated according to Eq.~\eqref{eqn:LensedFields} and Eq.~\eqref{eqn:AlphaRotation}. A realization map of instrument noise is then added to get $(\obsmap{Q}, \obsmap{U})$. 
We focus on two experiments: one that is noiseless and another that is comparable to the noise proposed for CMB-S4~\cite{Abazajian:2016yjj,CMB-S4:2020lpa} ($\Delta_{T}=1.0~\mu$K-arcmin, $\theta_{\mathrm{FWHM}}=1.4'$). 
Similar to what was done in Refs.~\cite{Guzman:2021nfk} and~\cite{Guzman:2021ygf}, when generating the CMB map data set, a randomly selected portion of CMB maps had $\kappa$ and $\alpha$ set to zero to create a 'null' set. No cosine taper is applied to the maps after their simulation. Additionally, we do not make any cuts on $\ell$ to the simulated maps during training or the subsequent analysis.

The map simulation process described above is repeated to generate a complete data set of \num{40000} training, \num{5000} validation, and three sets of \num{5000} prediction images for a total of \num{60000} sets of maps.  We simulated \num{60000} maps for each of the two noise levels we consider. For each noise level we trained a single separate neural network as well. The prediction data sets do not include maps from the null set (those with $\kappa=\alpha=0$). Furthermore, depending on the purpose of the data set, the CMB maps are generated with tensor-to-scalar ratios of $r \in (0, 0.1, 0.001, 0.0001)$. For the training and validation data sets, we use a value of $r=0$. The three prediction data sets used for analysis are each generated with a specific $r$ value of \num{0.1}, \num{0.001}, and \num{0.0001}. More information about the motivation behind the choices of $r$ will be provided in Sec.~\ref{sec:Results}.

\subsection{Network Training}
In total, two ResUNet-CMB networks were trained -- one for each noise level. Each ResUNet-CMB network was trained using a multi-GPU computer cluster consisting of NVIDIA A100 GPUs. To increase training speed, data parallelization was implemented with \texttt{Horovod}\footnote{\url{https://github.com/horovod/horovod}\\}~\cite{sergeev2018horovod}. In the data parallelization method we employed, a copy of the network is made to each individual GPU with the randomly initialized neural network weights broadcast to all copies. Each copy of the network is then sent a batch of the training data on which it performs operations. After all training data has been processed in one training epoch, the gradients calculated by each copy of the network are collected, averaged together, and used to update the weights of a single network copy. The updated weights are then broadcast to all network copies. As a result of the training process, the effective batch size per epoch increases. Thus, the learning rate must also be scaled to accommodate the larger effective batch size~\cite{sergeev2018horovod, mlbook}.

For training, we use a total of \num{16} A100 GPUs. We choose to use a scaled learning rate of $0.25/16$ with 8 warm-up epochs~\cite{warmupEpoch}. We follow our previous work and use the Adam optimizer and set a constant dropout rate of \num{0.3} for all dropout layers. We also decay the learning rate by a factor of \num{0.5} if after three training epochs there has been no decrease in validation loss. Training is halted and the model with the lowest validation loss is saved if there has been no decrease in validation loss after \num{25} training epochs. All other training hyperparameters match those in our previous work~\cite{Guzman:2021nfk, Guzman:2021ygf}. On average, training took around \num{2} hours for the noiseless case and \num{1} hour for the case with noise. 

While we choose to train the network using a large allocation of computational resources to reduce training time, this network with the larger image size data can still be trained on the more accessible graphics cards and resources used in Refs.~\cite{Guzman:2021nfk, Guzman:2021ygf} if modifications to training hyperparameters are made. With a prediction batch size of \num{32}, making a single prediction of ($\kappa$, $\alpha$) from a single set of ($Q$, $U$) polarization maps takes approximately \num{0.02}~seconds on an NVIDIA P100 GPU.

\begin{figure*}[t!]
    \centering
    \includegraphics[width=\textwidth]{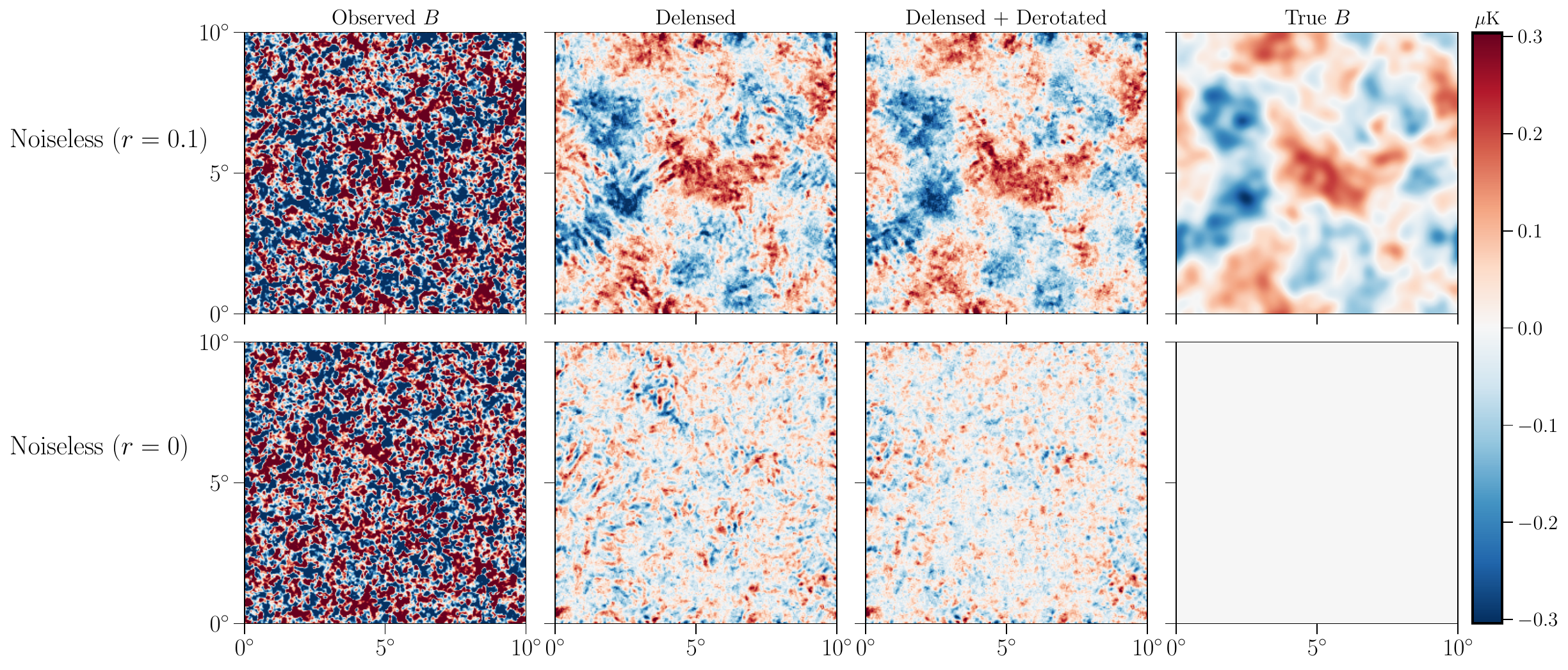}
    \caption{
    Example $B$-mode polarization maps showing the progression of going from observed to delensed then derotated and delensed CMB maps. Delensing and derotation are performed on $(\obsmap{Q}, \obsmap{U})$ maps and converted to $B$ polarization for each step in this figure. One can see that the amplitude of the residual $B$-modes are reduced after an application of delensing and even further minimized after derotation. The effect of derotation is more visible in the polarization map with $r=0$.
    }
    \label{fig:map_results_pipeline} 
\end{figure*}

\section{Results}
\label{sec:Results}
In this section, we present results for residual B-mode signal detection after delensing and derotation has been applied with the $\kappa$ and $\alpha$ predictions from the fully trained ResUNet-CMB network. We show map, spectra, and $r$ distribution results for two experimental noise levels. We will show that the results match well with what would be expected from an optimal iterative analysis similar to what is described in Sec.~\ref{sec:SecondaryAnisotropies}.

As mentioned in Refs.~\cite{Guzman:2021nfk},~\cite{Guzman:2021ygf}, and~\cite{Caldeira:2018ojb}, the neural network predicted distortion maps have a multiplicative bias. As we will show, inferring the tensor-to-scalar ratio for a set of $(\obsmap{Q}, \obsmap{U})$ maps can be achieved while using the biased $(\biasmap{\kappa}, \biasmap{\alpha})$ for the steps of delensing and derotation. For the purpose of delensing and deroation, the biased maps can be treated as if they are unbiased maps that have been Wiener filtered to down-weight the noise. The multiplicative map bias correction from Ref.~\cite{Guzman:2021ygf} is not required and is not applied for this analysis.

The analysis pipeline is as follows. We first take the $(\obsmap{Q}, \obsmap{U})$ maps in the prediction data set, pre-process them according to what is outlined in Sec.~\ref{sec:DataPipeNetTraining}, and make predictions of $\biasmap{\kappa}$ and $\biasmap{\alpha}$ using the trained ResUNet-CMB network. Thus, \num{5000} sets of $(\obsmap{Q}, \obsmap{U})$ result in \num{5000} sets of $(\biasmap{\kappa}, \biasmap{\alpha})$ neural network predictions. Next, we take the same $(\obsmap{Q}, \obsmap{U})$ we used as input to the trained neural network and take the corresponding predicted set of post-processed $(\biasmap{\kappa}, \biasmap{\alpha})$ and delens, then derotate the observed CMB polarization maps.  

We found that the best results were obtained by simply directly using the maps $(\biasmap{\kappa}, \biasmap{\alpha})$ for delensing and derotation, with no filtering applied to these maps.  As we showed in Refs.~\cite{Guzman:2021nfk,Guzman:2021ygf}, the reconstructed maps produced by ResUNet-CMB behave as if they have already been Wiener filtered, since the network predicts vanishing contributions to the maps on the scales that are poorly reconstructed.  It is possible to correct the multiplicative bias and estimate the noise power for the reconstructed maps (as was done in Refs.~\cite{Guzman:2021nfk,Guzman:2021ygf}).  We could then Wiener filter the unbiased maps; however, we found that this procedure results in maps that very closely resemble $(\biasmap{\kappa}, \biasmap{\alpha})$ with slightly larger scatter. We therefore employ $(\biasmap{\kappa}, \biasmap{\alpha})$ directly rather than explicitly constructing $W\star\hat{\phi}$ and $W\star\hat{\alpha}$.

In the data pipeline outlined in Sec.~\ref{sec:DataPipeNetTraining}, the CMB maps were lensed with \texttt{Orphics} which uses \texttt{Pixell}\footnote{\url{https://github.com/simonsobs/pixell}} as a backend. To delens, \texttt{Pixell} employs an iterative delensing scheme that is equal to the inverse of the lensing procedure. First, the lensing convergence is converted to the lensing potential, $\phi$. The deflection field, $\nabla \phi$, is calculated which results in the coordinate displacements in pixel space. Then, the inverse gradient is found through an iterative procedure. As a result, the scheme used by \texttt{Pixell} is all-orders delensing. After delensing the $(\obsmap{Q}, \obsmap{U})$ maps with $\biasmap{\kappa}$, we then derotate the observed maps by applying Eq.~\eqref{eqn:AlphaRotation} with $\alpha=-\biasmap{\alpha}$.  Our procedure is therefore an implementation of what is shown in Eq.~\eqref{eq:Delensed_Derotated_Map}, with the outputs of ResUNet-CMB $(\biasmap{\kappa}, \biasmap{\alpha})$ used in place of $W\star\hat{\phi}$ and $W\star\hat{\alpha}$. Throughout this section, we will discuss and show results from the steps of delensing and after delensing and derotation has been applied.

\subsection{Map-Level Results}
\label{sec:Map_Results}

We begin by presenting the map level results which provide an illustration of the secondary removal pipeline for a single realization. In Fig.~\ref{fig:map_results_pipeline}, we show the secondary removal steps at map level for the noiseless experiment for two different values of $r$. In the top panels, we show simulated maps with a relatively large value of $r$ to demonstrate a regime where signal would dominate over noise for a significant detection.
When going from observed to delensed, a large part of the $B$-mode power induced by lensing is removed. The underlying large scale structure dominated by primordial contributions is seen and faithfully reconstructed compared to the truth $B$-mode polarization. Residual lensing and derotation induced $B$-mode polarization power is still visible on small angular scales. After derotating with the neural network predicted $\alpha$ map, the induced $B$-modes are further reduced, giving a result that more closely matches the true primordial $B$-mode polarization map. In the bottom panels, we show a simulation we label as 'null' which is where the tensor-to-scalar ratio is set to zero. For the null experiment, all of the $B$-mode polarization present in the map arises as a result of the secondary sources conversion from $E$ polarization. After delensing, the reduction in induced $B$-mode polarization matches what is seen in the corresponding map shown in the top panel. After derotation, the induced $B$ polarization removed by the operation is more easily seen in the null experiment.

\subsection{Spectrum-Level Results}
\label{sec:Spectrum_Results}

\begin{figure}[tbp!]
    \centering
    \includegraphics[width=\linewidth]{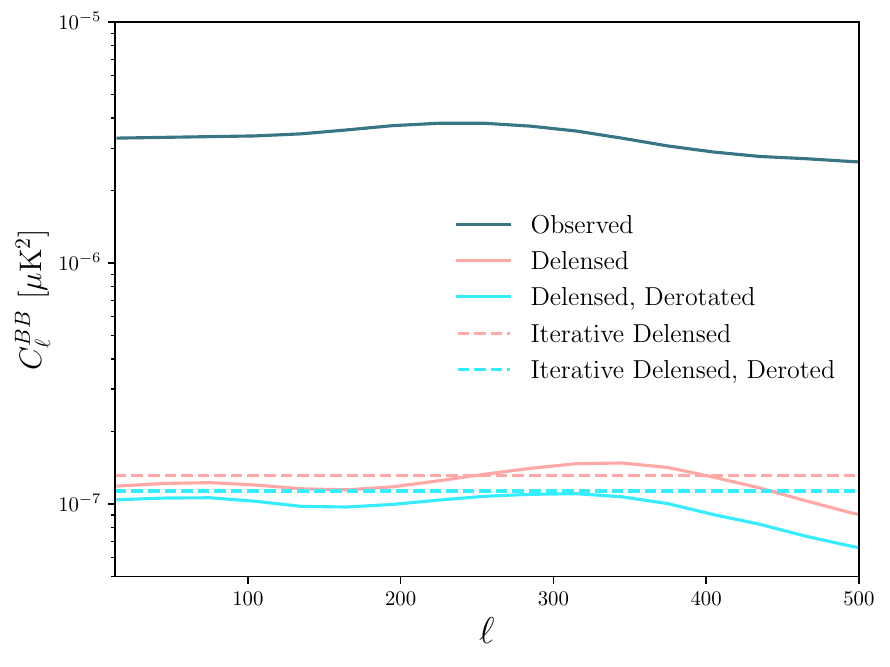}
    \caption{
    Averaged power spectra after three different steps in the secondary removal process (solid lines) and the estimated power from an ideal iterative procedure (dashed). Power spectra (solid lines) are calculated by averaging over \num{5000} simulated CMB maps of the noiseless prediction data set.
    }
    \label{fig:null_b_ps}
\end{figure}

\begin{figure}[tbp!]
    \centering
    \includegraphics[width=\linewidth]{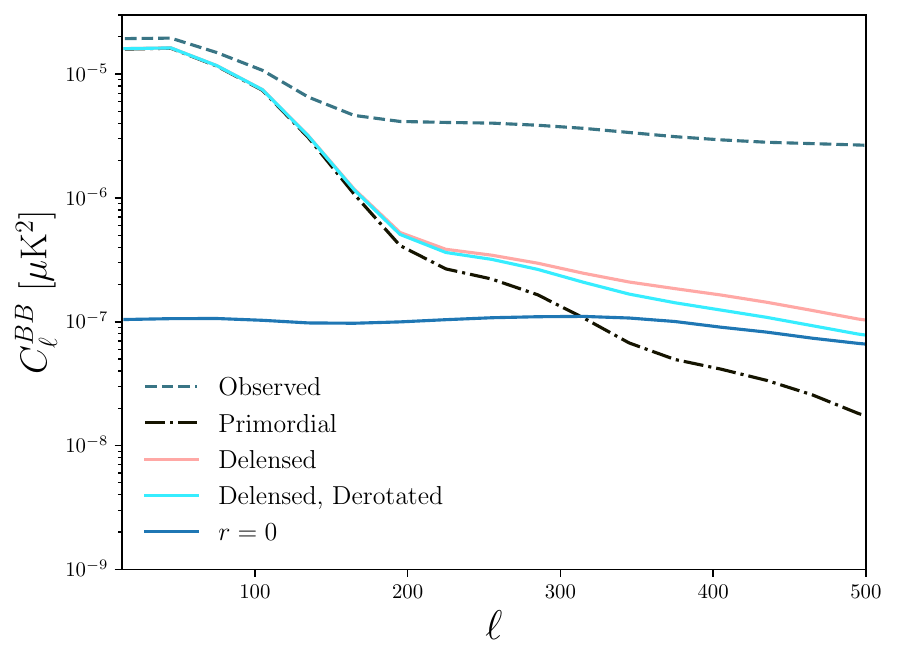}
    \caption{
    Averaged power spectra of B-mode polarization maps for every stage of secondary anisotropy removal compared against the primordial and null ($r=0$) power spectra. Power spectra are from simulated CMB maps (prediction data set) with a tensor-to-scalar ratio of $r=0.1$. Power spectra are averaged over \num{5000} simulated CMB maps from the noiseless experiment. The $r=0$ spectrum is equivalent to the delensed and derotated spectrum from Fig.~\ref{fig:null_b_ps} and is included for easing comparisons between the two figures.
    }
    \label{fig:primb_ps}
\end{figure}

We now further quantify the results by showing the recovery of the primordial $B$-mode polarization power spectrum. To calculate the power spectra shown in the figures, we first calculate the power spectrum of each individual recovered $B$-mode polarization map in the prediction data set, $C_{\ell}^{BB}$. The result is then averaged to get $\langle C_{\ell}^{BB} \rangle$, which we then plot for comparison. In Fig.~\ref{fig:null_b_ps}, we show the average power spectra at different stages of secondary anisotropy removal with $r=0$ for the noiseless experiment. The spectra mirror what is observed in the simulated maps in Fig.~\ref{fig:map_results_pipeline} where there is a larger reduction in the residual induced $B$-mode power after delensing the observations and a smaller decrease after derotation. The dashed lines in the figure represent an estimate from an ideal iterative technique described in Sec.~\ref{sec:SecondaryAnisotropies}.  For the entire $\ell$ range we consider, the ResUNet-CMB delensed spectrum, and the delensed and derotated spectrum approach those of the iterative method to within approximately $10\%$ percent.

Figure~\ref{fig:primb_ps} shows the recovered B-mode polarization power spectrum for a tensor-to-scalar ratio of $r=0.1$ for the noiseless experiment. For this figure, the primordial power spectrum is calculated as the average power spectrum of the primordial (truth) $B$ maps from the corresponding prediction data set. At large angular scales, $\ell<150$, the delensing operation recovers the primordial $B$ signal faithfully. For larger values of $\ell$ (smaller angular scales), the improvement in $B$-mode polarization power recovery due to derotation becomes evident. Figures~\ref{fig:null_b_ps} and \ref{fig:primb_ps} convey similar information with the key difference being the value of $r$ used for the spectra. The $r=0$ curve in Fig.~\ref{fig:primb_ps} is equivalent to the delensed and derotated line in Fig.~\ref{fig:null_b_ps} which we plot to ease comparison between the two figures.

\subsection{Tensor-to-Scalar Ratio Estimates}
\label{sec:r_Results}

Finally, we use the recovered $B$-mode polarization maps to estimate the tensor-to-scalar ratio $r$. We also compare how performing delensing and derotation with the neural network predicted $\biasmap{\kappa}$ and $\biasmap{\alpha}$ maps affects the inferred $r$ values and the associated confidence intervals of the inference.

To infer the value of $r$ from a recovered $B$-mode polarization map, we find the maximum value of a modified version of the likelihood described in  Ref.~\cite{Hamimeche:2008ai},
\begin{equation}\label{eqn:rLikelihood}
    \log{\mathcal{L}} = -\frac{1}{2} \sum_{\ell \ell'} g_\ell C_{\ell,\mathrm{fid}} [M^{-1}_{\mathrm{fid}}]_{\ell \ell'} C_{\ell,\mathrm{fid}} g_{\ell'} \, ,
\end{equation}
with $g_\ell$ defined as,
\begin{equation}\label{eqn:likegdef}
    g_\ell = \sqrt{2(V_\ell-\ln(V_\ell)-1)} \, ,
\end{equation}
and $V_\ell$ taken to be,
\begin{equation}\label{eqn:likeVdef}
    V_\ell = \frac{C_\ell^{BB, \mathrm{recov}}}{\langle C_\ell^{BB, \mathrm{null}}\rangle + \langle C_\ell^{BB, \mathrm{prim}, \mathrm{ref}}\rangle \frac{r}{r_{\mathrm{ref}}}} \, ,
\end{equation}
where we have used the fact that the primordial $B$-mode power spectrum scales linearly with the tensor-to-scalar ratio, $C_\ell^{BB,\mathrm{prim}}\propto r$.
In Eq.~\eqref{eqn:rLikelihood}, the 'fid' represents a fiducial model which we choose to be the average power spectrum for the null case of $r=0$ which is $\langle C_{\ell}^{BB}\rangle_{r=0}$. The angled brackets, $\langle \ldots \rangle$, represent the average over the prediction data set. The covariance of the fiducial model is given by $M$, which we measured from the prediction data set and applied a Hartlap correction~\cite{Hartlap:2006kj}. In Eq.~\eqref{eqn:likeVdef}, 'recov' refers to delensed and derotated spectra and 'null' is the same as defined above, when $r=0$. The $r_{\mathrm{ref}}$ value is the value of the tensor-to-scalar ratio used for the primordial spectrum of the prediction data set we are using.

\begin{figure}[tbp!]
    \centering
    \includegraphics[width=\linewidth]{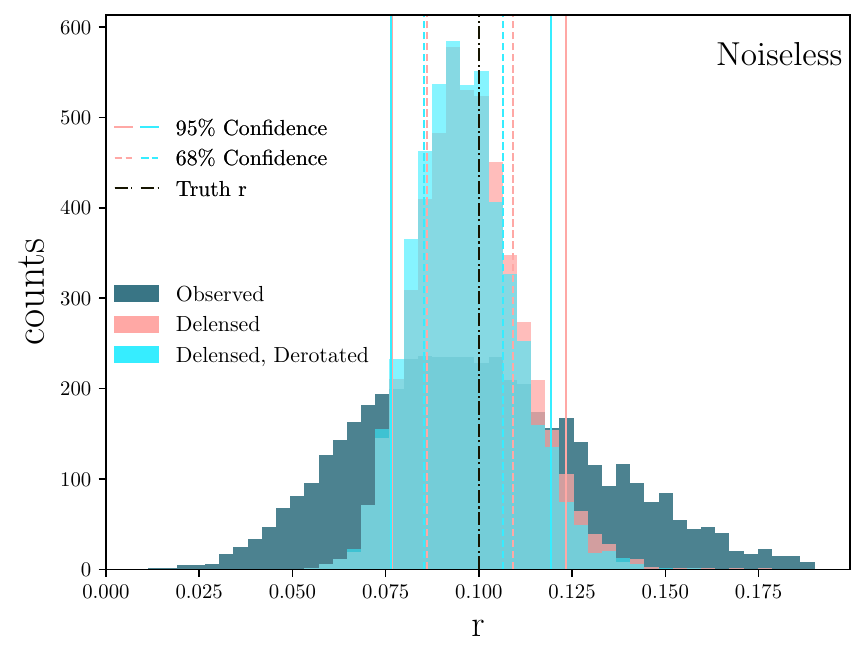}
    \caption{
    Histogram of all inferred $r$ values ($r=0.1$) from the \num{5000} likelihood predictions obtained from the noiseless data set shown at two stages of secondary removal: after delensing in coral and after delensing and derotation in cyan.  Delensing provides significantly tighter constraints and derotation subsequently provides an additional small improvement.  The estimates remain unbiased after each step in the procedure. 
    }
    \label{fig:histogramB_1em1}
\end{figure}

\begin{figure}[tbp!]
    \centering
    \includegraphics[width=\linewidth]{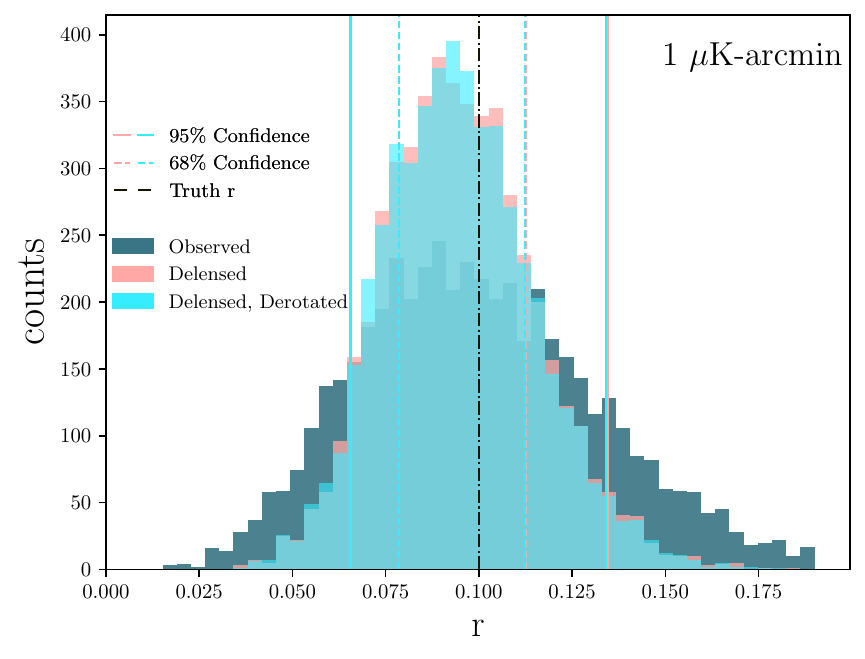}
    \caption{
    Same as Fig.~\ref{fig:histogramB_1em1}, but for the simulations with $\Delta_{T}=1~\mu$K-arcmin noise.
    }
    \label{fig:histogramB_1em1_1uk}
\end{figure}

\begin{figure}[tbp!]
    \centering
    \includegraphics[width=\linewidth]{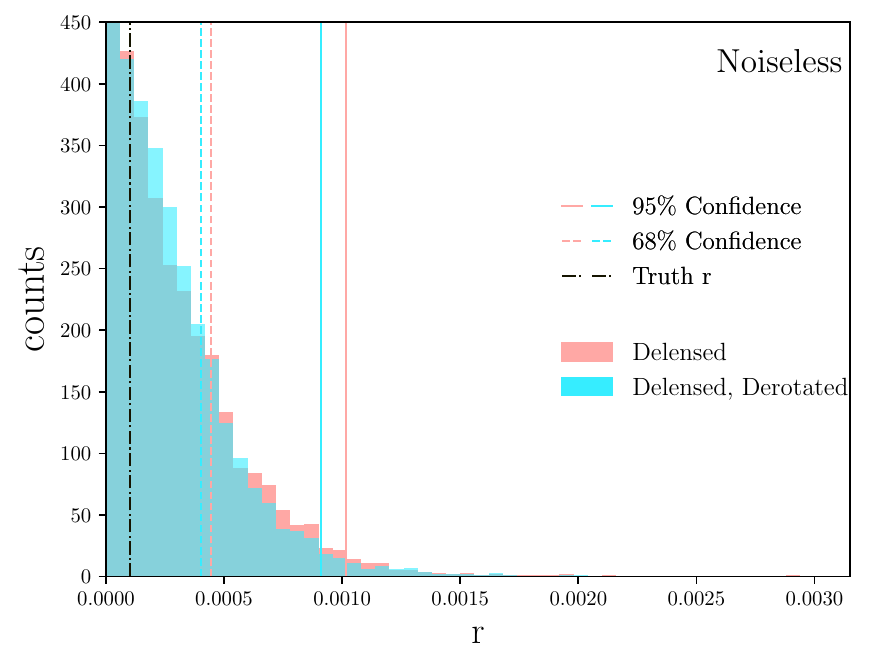}
    \caption{
    Same as Fig.~\ref{fig:histogramB_1em1}, but for the simulations with $r=0.0001$.  Here we do not show the predictions applied to the observed maps before delensing, since the distribution is too wide to be neatly contained in the plot.
    }
    \label{fig:histogramB_1em4}
\end{figure}

\begin{figure}[tbp!]
    \centering
    \includegraphics[width=\linewidth]{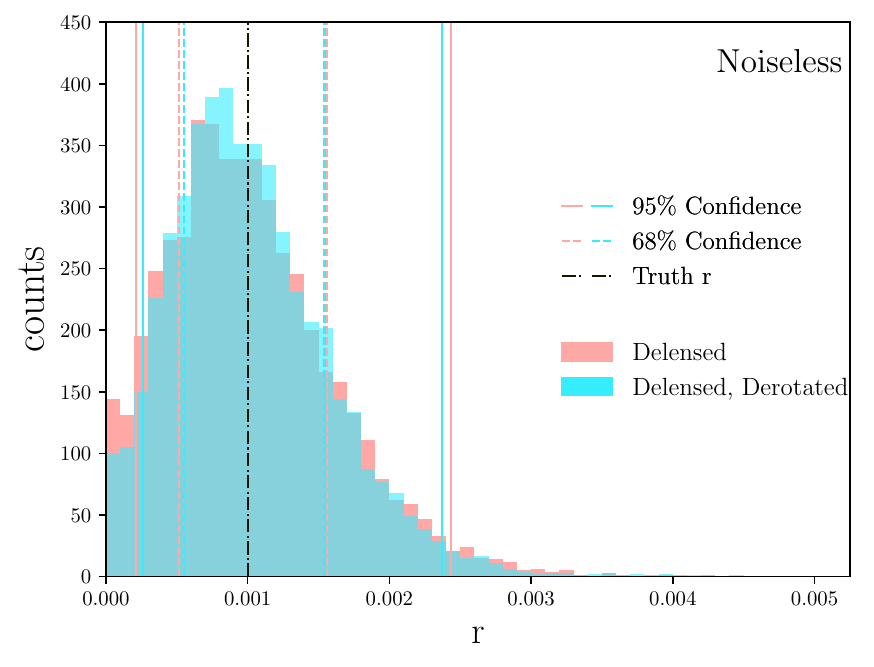}
    \caption{
    Same as Fig.~\ref{fig:histogramB_1em4}, but for the simulations with $r=0.001$.
    }
    \label{fig:histogramB_1em3}
\end{figure}

After finding the maximum likelihood value, $r_{\mathrm{pred}}$, we then calculate the $68\%$ and $95\%$ confidence intervals for a particular likelihood by finding $r_{\mathrm{lower}}$ and $r_{\mathrm{upper}}$, which are the lower and upper bounds on the confidence. For our calculations, $r_{\mathrm{lower}}$ and $r_{\mathrm{upper}}$ are found by ensuring the following equation is satisfied for a specific confidence interval (C.I.),
\begin{equation}\label{eqn:confidenceInterval}
    \frac{\int_{r_{\mathrm{lower}}}^{r_{\mathrm{upper}}} \mathcal{L}(r) dr}{\int_{r_{\mathrm{min}}}^{r_{\mathrm{max}}} \mathcal{L}(r) dr} = C.I. \, ,
\end{equation}
where $r_{\mathrm{max}}$ and $r_{\mathrm{min}}$ are the max and minimum $r$ values included in the likelihood curve produced. Eqs.~\eqref{eqn:rLikelihood} and~\eqref{eqn:confidenceInterval} are repeated for each recovered $B$-mode polarization map in the prediction data set to generate \num{5000} $r$ values and \num{5000} upper- and lower-confidence bounds for each.

The results for the tensor-to-scalar ratio of $r=0.1$ for the noiseless experiment are shown in Fig.~\ref{fig:histogramB_1em1}. The histogram shows the distribution of the maximum likelihood $r_{\mathrm{pred}}$ values for three different steps in the secondary removal process. The confidence intervals are only shown for the delensed and the delensed and derotated operations. The final confidence intervals are calculated as median values of the \num{5000} individual likelihood confidence intervals. The observed histogram is found by applying Eqs.~\eqref{eqn:rLikelihood} and~\eqref{eqn:confidenceInterval} to the observed $B$-mode polarization prediction maps. After delensing, there is a significant decrease in the variance of the $r$ distribution. There is a further slight decrease in variance after further performing derotation on the delensed maps. It can be seen that the lower confidence bounds are nearly identical after delensing and after further derotating the maps. However, the upper bounds see a slight improvement at both the $68\%$ and $95\%$ confidence level after the derotation operation. All three histograms are peaked closely around the expected fiducial $r$ value of $0.1$. This result shows an unbiased and reduced variance in $r_{\mathrm{pred}}$ as a result of both the delensing and derotation operations.\footnote{All of our estimates, whether or not delensing and derotation are applied, appear to show a small bias toward a very slight under-prediction of the true value of $r$, at a level well below a 1-$\sigma$ shift.  This small offset may be due to a minor deficiency in our simulations or in the likelihood we employ, but since it appears even in cases where no delensing and derotation are applied, we do not pursue it further here. }

To offer a comparison to the noiseless experiment results, in Fig.~\ref{fig:histogramB_1em1_1uk} we show the histograms for $r=0.1$ from the $1~\mu$K-arcmin noise experiment. As expected, while all three histograms peak around the true $r$ value, the variance is much larger than in the noiseless experiment. Delensing produces bounds on the inferred value of $r$ that are $40\%$ tighter than the the case of no delensing, while the improvement from derotation is negligible in this case. However, this result shows that with just a naive application of the $\biasmap{\kappa}$ and $\biasmap{\alpha}$ in the secondary removal process with no extra bias corrections or Wiener filtering, we are able to produce the expected outcome.

Figure~\ref{fig:histogramB_1em4} shows the distribution for a fiducial value of $r=0.0001$ for the noiseless experiment.
In this figure, we do not show the observed histogram since its variance would be large enough where it would obscure the other distributions. Furthermore, only the upper confidence bounds are shown as the lower are equivalent to zero. When going from observed to delensed, the standard deviation of the distribution decreases by around $95\%$. After derotation of the delensed maps, the standard deviation sees a further decrease by approximately $7\%$. The upper confidence bounds both see an improvement from the delensed confidence bounds.

So far, we have shown two of the more extreme values for $r$. In Fig.~\ref{fig:histogramB_1em3} we plot the distribution for the noiseless experiment with $r=0.001$. From the figure, we see that both the delensed and the delensed and derotated distributions peak around the true primordial $r$ value. The delensed and derotated distribution has improved upper and lower confidence bounds. When the value of fiducial $r$ is between the largest and smallest values we consider, we are able to place tighter constraints on $r$ after delensing and derotation.

All the results in this section have shown that using the ResUNet-CMB predictions, $\biasmap{\kappa}$ and $\biasmap{\alpha}$, for the delensing and derotation of the observed polarization maps results in an inferred $r$ distribution that is unbiased and has a reduced variance.

\section{Conclusion}
\label{sec:Conclusion}
In this paper, we showed that the simultaneous distortion field estimates made from the ResUNet-CMB neural network can be used in a complete analysis pipeline to infer the value of $r$ and produce improved confidence bounds with a variance that matches an estimate from an idealized iterative method.
We used the ResUNet-CMB network described in Ref.~\cite{Guzman:2021nfk} to simultaneously predict the anisotropic cosmic polarization rotation field $\alpha$ and the lensing convergence $\kappa$ from CMB maps ($\obsmap{Q}$, $\obsmap{U}$) simulated with a nonzero tensor-to-scalar ratio. We then used the ResUNet-CMB estimates with no bias corrections to perform delensing and derotation on the observed polarization maps to remove the secondary anisotropies and recover the primordial $B$-mode polarization signal. We showed that after a simple application of the inverse polarization rotation operation with the estimated $\alpha$ and all-orders delensing with the estimated $\kappa$, the secondary $B$-mode polarization signal was significantly reduced. This procedure was shown to work without modification on maps containing various levels of primordial $B$-mode power.
We also showed that the inference of $r$ derived from the recovered $B$-mode polarization map had improved upper limit constraints for the lowest value of $r$ examined after derotation was applied to the delensed maps. Furthermore, for intermediate values of the fiducial $r$, tighter constraints on the parameter value can be made. Finally, we found that delensing and derotating with the ResUNet-CMB estimates did not require the map bias correction used in previous works~\cite{Caldeira:2018ojb,Guzman:2021nfk, Guzman:2021ygf, Heinrich:2022hsf}. 

The neural network in Ref.~\cite{Heinrich:2022hsf} was trained with simulated maps with varying values of $r$. They found the inclusion of nonzero $r$ values did not harm the lensing reconstruction made by the convolutional neural network. In this work, we chose to train the network on simulated CMB maps where $r=0$ and instead make predictions on maps with nonzero $r$. With this method we found that we are able to achieve a nearly optimal reconstruction and the subsequent delensing and derotation of the observed polarization maps also matched the estimated optimal result. Furthermore, in Ref.~\cite{Heinrich:2022hsf}, template delensing is performed to recover the primordial $B$-mode polarization map. In template delensing, the observed $E$-mode polarization and $\hat{\phi}$ estimate are used to construct a template, or map, of the lensing induced $B$-mode polarization. The template $B$-mode polarization is then subtracted from the observed $B$-modes to get the delensed $B$-mode map. However, it has been shown that template delensing is sub-optimal compared to the all-orders delensing method we use~\cite{BaleatoLizancos:2020jby}. The use of template delensing also necessitated changes to the training method of the neural network in Ref.~\cite{Heinrich:2022hsf}.
In this work, we found that all-orders delensing could be done directly with the biased estimate maps to achieve a nearly optimal result without the need for $\ell$ cuts on the CMB maps.

In our previous works~\cite{Guzman:2021nfk, Guzman:2021ygf}, we found the ResUNet-CMB network faithfully reconstructed multiple distortion fields simultaneously with minimal changes to network architecture which mitigated the addition of computational complexity. In this work, we further show the extensible nature of the ResUNet-CMB network by using CMB maps that are twice as large as those in our previous work. Since we increased the image size without changing the internal layer structure and parameters of the network, the receptive field covered a smaller region of the input image. However, despite the smaller receptive field, the network continued to make optimal reconstructions. Additionally, the larger image size moves us a step closer to creating a network capable of handling the image sizes used by CMB surveys.

While we have shown that the ResUNet-CMB estimates can be used to great success for the removal of secondary anisotropies in a full analysis pipeline for inferring the tensor-to-scalar ratio of observed CMB polarization maps, there are more areas to explore in order to gain a full understanding of the potential impact machine learning can have on this procedure.
We used images with a size of $256\times256$ pixels with no modifications made to the depth or inner structure of the neural network. One possible improvement would be to explore deeper and wider network architectures to see if further gains to the result can be obtained. Increasing the effective receptive field and adding model complexity may prove to be beneficial for larger images.

In this work, we found the tensor-to-scalar ratio and confidence interval with Bayesian inference. Another possible extension to this work would be to alter the ResUNet-CMB network to also output its own estimate of $r$ and compare how well it matches compared to the Bayesian inference method. Furthermore, recent advances in uncertainty quantification methods for machine learning applications~\cite{Kuestner2024DeepEnsembleUQ, He:2023abc,fischer2023uc,ekmekci2025} may provide additional strategies for inference, including devising a neural network that produces both an estimate of the $r$ value as well as its uncertainty.

In addition to improved inference of the primordial gravitational wave amplitude, it has been shown that delensing can provide improved constraints on a wide range of other cosmological parameters, through the impact that delensing has on the temperature and $E$-mode polarization spectra~\cite{Green:2016cjr,Hotinli:2021umk,Ange:2023ygk,Trendafilova:2023xtq}. The delensing and derotation pipeline described here could be applied to both temperature and polarization maps to take advantage of this potential.

We demonstrated that the outputs from ResUNet-CMB can be used in a full analysis pipeline to produce improved estimates of the  amplitude of primordial gravitational waves in simulated CMB maps. This success gives strong motivation to continue extending the ResUNet-CMB network and other machine learning approaches to tease out signatures of fundamental physics in CMB surveys and beyond.


\vspace*{2em}

\section*{Acknowledgments}
We thank Alexander van Engelen and Jasmine Liu for helpful conversations.
This work is supported by the US Department of Energy Office of Science under grant no.~DE-SC0010129, by NASA through ADAP grant \mbox{80NSSC24K0665}, and by NSF through grant \mbox{AST-2510926}. EG was also supported by a Southern Methodist University Computational Science and Engineering fellowship. 
Computational resources for this research were provided by SMU’s O’Donnell Data Science and Research Computing Institute.

\bibliographystyle{utphys}
\bibliography{refs}

\end{document}